\pdfoutput=1 
\documentclass[aps,prc,floatfix,superscriptaddress,preprintnumbers,
twocolumn
]{revtex4-1}

\usepackage{graphicx,subfigure,latexsym}
\usepackage{amsmath,amssymb,amsfonts}
\usepackage[normalem]{ulem}
\usepackage{color}
\usepackage{bm}
\usepackage{url}
\usepackage[colorlinks,pdfstartview=FitH]{hyperref}
\hypersetup{linkcolor=blue,citecolor=blue,filecolor=black,urlcolor=blue}
\usepackage{comment}

\graphicspath{{plots/}}

\usepackage{custom-macros}

\begin{document}

\title{Reversing a heavy-ion collision
}

\preprint{EFI-14-9}

\author{Mikhail Stephanov}
\affiliation{
Physics Department,  University of Illinois at Chicago,
Chicago, IL, 60607%
}%
\affiliation{
 Enrico Fermi Institute, University of Chicago,
Chicago, IL 60637}

\author{Yi Yin}
\affiliation{
Physics Department,  University of Illinois at Chicago,
Chicago, IL, 60607%
}

\date{\today}


\begin{abstract}
  We introduce a novel approach to study the longitudinal hydrodynamic
  expansion of the quark-gluon fluid created in heavy-ion
  collisions. It consists of two steps: First, we apply the {\em
    maximum entropy method} to reconstruct the freeze-out surface from
  experimentally measured particle distribution.  We then take the output
  of the reconstruction as the ``initial'' condition to evolve the
  system {\em back} in time by solving the $1+1$ ideal hydrodynamic
  equations {\em analytically}, using the method of Khalatnikov and
  Landau.  We find an approximate Bjorken-like plateau in the energy
  density vs rapidity profile at the early times, which shrinks with
  time as the boundary shocks propagate inward.  In Bjorken frame, the
  fluid velocity is close to zero within the plateau, as in the Bjorken
  solution, but increases outside the plateau. The results carry
  implications for fully numerical hydrodynamic simulations as well
  as models of heavy-ion collisions based on gauge-gravity
  duality.

\end{abstract}


\maketitle

\section{Introduction}
\label{sec:introduction}
In 1983, Bjorken proposed to describe the central region of the
ultra-relativistic heavy-ion collisions by a
boost-invariant solution of the hydrodynamic
equations~\cite{Bjorken:1982qr}. In this elegant solution,
the fluid remains homogeneous and at rest in the Bjorken's
coordinates. The expansion of the system is entirely encoded in the
expanding nature of the hyperbolic Bjorken coordinates, very similar
to the Hubble expansion, but in one dimension.  A direct implication of
such boost-invariance at late times would be a mid-rapidity plateau in
the particle production.  Such a plateau is not apparent at
RHIC~\cite{Bearden:2004yx,Back:2004je} or
LHC~\cite{Abbas:2013bpa} (see Fig.~\ref{fig:dNdY} below).  Experimental
results suggest that it is imperative to study longitudinal expansion
beyond boost-invariant Bjorken model.

Fully numerical $1+1$~\cite{Satarov:2006iw,Bozek:2009ty} and
$3+1$~\cite{Hirano:2001yi,Schenke:2010nt} relativistic hydrodynamic
simulations 
{are appropriate tools} to address this challenge.
Unlike the highly-symmetric idealized Bjorken model, in which the unknown initial
condition is characterized by a single number (energy density, or
initial time), a more realistic simulation requires much more
information about the initial density and flow profile. While the
initial profile of the fireball in the {\em transverse plane} has been
studied extensively, our knowledge in longitudinal direction is
comparatively poor.  In practice, all the above simulations
\cite{Hirano:2001yi,Satarov:2006iw,Bozek:2009ty,Schenke:2010nt},
rely on an ansatz with several free parameters for the initial profile.
Those free parameters are tuned to match multiplicity distributions
obtained as a result of the simulations to the
experimental data.  This ``trial and error" method of
determining the initial conditions is clearly computationally intensive.

In this paper 
we introduce a novel approach to 
determine the
 longitudinal hydrodynamic
flow profile of the fireball created in heavy-ion collisions. 
It features two main
ingredients which distinguishes it from the traditional
approaches. First, we use the experimental data directly as an input to determine
the hydrodynamic variables on the
freeze-out surface. Second, we evolve the system {\em back} in time
using the freeze-out as ``initial'' condition. We are able to perform the
evolution {\em analytically}.

Correspondingly, our approach involves two key steps.  
First, we apply
the maximum entropy method (MEM)~\cite{wu1997maximum,bryan} to invert
Cooper-Frye freeze-out~\cite{Cooper:1974mv} prescription relating the
hydrodynamic variables at the freeze-out hypersurface to the
rapidity-dependent hadron spectrum measured in the experiments.  
As a state-of-the-art deconvolution technique, the maximum entropy method
(MEM) has proven to be a powerful tool in other branches of
physics~\cite{Jarrell1996133,wu1997maximum}, 
including lattice QCD~\cite{Asakawa:2000tr} and cosmology~\cite{Bajkova:1992}.  
To the best of our knowledge, 
however,
this is the first time that MEM has been applied to
the physics of heavy-ion collisions.

In the second step, we take advantage of the fact that $1+1$ ideal hydrodynamic
equations with initial conditions given on the isothermal surface can
be solved analytically.  Since the ideal hydrodynamic equations are
deterministic, we can evolve the system from freeze-out temperature
to early times with the information of freeze-out surface we obtained
in the first step.  As a result, we are able to reconstruct the history of
the longitudinal expansion and infer the longitudinal profile of the fireball at
early times.

We find that, while the observed particle distribution does not
show a rapidity plateau, the reconstructed hydrodynamic flow did have 
a rapidity plateau at earlier times. The width of the plateau shrinks
with time as the boundary shocks propagate towards the center of the fireball.

This paper is organized as follows.  In Sec.~\ref{sec:freeze_out}, we
discuss the deconvolution of Cooper-Frye freeze-out using the maximum
entropy method.  In Sec.~\ref{sec:hydro_solution}, we review the
general solution of $1+1$ ideal hydrodynamics.  Our results are
presented in Sec.~\ref{sec:results}.  We show the reconstructed
freeze-out surface obtained in the first step in
subsection~\ref{sec:surface}.  We then show the history of
longitudinal expansion in subsection~\ref{sec:longitudinalevolution}.
We compare the resulting early-time profile with that used in
hydrodynamic simulations.  Finally, in Sec.~\ref{sec:conclusion}, we
conclude.

\section{Reconstruction of freeze-out surface and maximum entropy method}
\label{sec:freeze_out}

We shall follow the standard Copper-Frye
approach~\cite{Cooper:1974mv} to relate the momentum-space
multiplicity distribution $d^3N_s/d^3 p$
 of the hadrons of a given species to the  temperature $T$ and
 flow velocity $u^\mu$ profiles on the freeze-out hyper-surface:
\begin{equation}
\label{eq:copper-Frye}
p^{0}\frac{d^3N_s}{d^3 p} = 
\frac{1}{\phase}\, p^{\mu} \int d^{3}\s_{\mu}(x)\, f(p; u(x), T(x)) \, ,
\end{equation}
where $f(p; u(x), T(x))$ is the local equilibrium particle
distribution function at point $x$ and $d^3\sigma_\mu(x)$ is the infinitesimal
volume 4-vector normal to the freeze-out hypersurface at this point.

In this exploratory analysis, we concentrate on the
longitudinal expansion and neglect, for simplicity, the effect of
transverse expansion.
We focus on the spectrum of pions -- most abundant hadron species. We 
neglect quantum statistics in the phase space distribution --
the generalization is straightforward.
In this case $f(p; u, T)$ is given by the Boltzmann
distribution longitudinally boosted by fluid rapidity $\alpha$:
\begin{equation}
f(p; u, T)
= e^{-\frac{p_{\mu}u^{\mu}}{T}}
= e^{-\frac{\mt}{T}\cosh(Y-\a)} \, ,
\end{equation}
where $\mt=\sqrt{\pt^2+m^2_{\pi}}$ is the pion transverse mass, $Y$ is the
particle rapidity in the lab frame, related to the particle 4-momentum as
\begin{equation}
  \label{eq:Ep}
(p^t,p^z)=m_\perp(\cosh Y,\sinh Y)
\end{equation}
 and $\a$ is the local fluid rapidity related
to flow 4-velocity by
\begin{equation}\label{eq:utuz}
\(u^{t},u^{z}\)=\(\cosh\a,\sinh\a\).
\end{equation}

We parameterize the freeze-out hyper-surface using Bjorken coordinates
$\tau$ and $\eta$
\begin{equation}\label{eq:taueta}
t= \tau\cosh\eta,\quad z=\tau\sinh\eta\,
\end{equation}
by expressing the freeze-out proper time as
a function of the Bjorken rapidity: $\tau_f(\eta)$. Using the
expression for the hyper-surface volume element vector in the Bjorken
coordinates:
\begin{equation}
(d^3\sigma_\tau,d^3\sigma_\eta)=\tau_fd\eta
d^2x_\perp(1,-\partial_\eta\tau_f(\eta))\label{eq:dsigma}
\end{equation}
we can write
\begin{equation}
p^\mu d^3\sigma_\mu
= m_\perp\partial_\eta\[\,\tau_f(\eta)\sinh(\eta-Y)\,\]\,d\eta
d^2x_\perp\,.\label{eq:p.dsigma}
\end{equation}

The local fluid rapidity
$\alpha$ is also a function of the Bjorken coordinate $\eta$. For example, purely 
Bjorken flow corresponds to $\alpha(\eta)=\eta$. We
shall assume that this function is monotonous, i.e., its
inverse, $\eta_f(\alpha)$, is single-valued.
The pion distribution in rapidity $Y$ and transverse momentum
$p_\perp$ now reads:
\begin{multline}
  \label{eq:Cooper-Frye-2d}
 \frac{d^3N}{ dYd^2\pt} 
 =\frac{ A_{\perp}\mt}{\phase} \int^{\infty}_{-\infty}d\a\,
 e^{-\frac{\mt}{\Tf}\cosh(Y-\a)} \\\times
\pd_{\a} \[\,\tau_f(\a)\sinh\(\eta_f(\a)-Y\)\,\]\, ,
\end{multline}
where $A_{\perp}$ is the total transverse area of the freeze-out
hypersurface.
We have
changed the integration 
variable parameterizing the freeze-out surface from $\eta$
to $\a$ and used $\tau_f(\a)$ as a shorthand for $\tau_f(\eta_f(\a))$.  
Integrating \eq\eqref{eq:Cooper-Frye-2d} by parts, we
can write
\begin{multline}
  \label{eq:Cooper-Frye-2dv2}
 \frac{d^3N}{dYd^2\pt} = 
\frac{\mt^2}{\phase \Tf} \int^{\infty}_{-\infty}d\a\, 
e^{-\frac{\mt}{\Tf}\cosh(Y-\a)}\,
\\\times
 \Big(\, 
  \sinh^2(Y-\a)\cdot [A_{\perp}\tau_f(\a)\cosh(\a-\eta_f(\a))] \\ +
  \sinh(Y-\a)\cosh(Y-\a)
 \cdot [A_{\perp}\tau_f(\a)\sinh(\a-\eta_f(\a))]
 \,\Big) \, ,
\end{multline} 
where we enclosed the factors carrying information about the freeze-out
surface (size, shape and flow velocity) in square brackets. These
functions of $\a$ will be important intermediate objects in our 
analysis, and we denote them as 
  \begin{align}
    &\rho_1(\a)\equiv
    A_{\perp}\tau_f(\a)\cosh(\a-\eta_f(\a))=A_{\perp}
    \tau_f(\a)u^{\tau}_f(\a)\,;\nonumber\\
    & \rho_2(\a)\equiv A_{\perp}\tau_f(\a)\sinh(\a-\eta_f(\a))=
    A_{\perp}\tau^2_f(\a)u^{\eta}_f(\a)\, \label{eq:rho12_def}
  \end{align}
where we also note that they can be expressed in terms of the
 components of fluid velocity in Bjorken coordinates:
\begin{equation}
\label{eq:utaueta}
u^{\tau}= \cosh\(\a-\eta\)\, , 
\qquad
u^{\eta}= \tau^{-1}\sinh\(\a-\eta\)\, ,
\end{equation}
evaluated on the freeze-out surface parameterized by $\alpha$: $\tau=\tau_f(\alpha)$, $\eta=\eta_f(\alpha)$.

The physical meaning of $\rho_{1,2}$ can be understood as follows.
The ratio
\begin{equation}
  \label{eq:veta-def}
  \rho_2/\rho_1=\tanh(\a-\eta_f(\a))\equiv v_\eta
\end{equation}
 is the
local flow velocity, while $\sqrt{\rho_1^2-\rho_2^2}=A_\perp\tau_f(\a)$ is
the volume per unit Bjorken rapidity both measured in the Bjorken
frame at a given point on the freeze-out hypersurface.

Our first task is to reconstruct the freeze-out surface, i.e., to
determine two independent functions: $\tau_f(\a)$ and
${\eta}_f(\alpha)$ appearing in Eq.~(\ref{eq:Cooper-Frye-2dv2}) by
matching Eq.~(\ref{eq:Cooper-Frye-2dv2}) to experimentally observed
particle spectrum.  The direct inversion of Cooper-Frye freeze-out
\eq\eqref{eq:Cooper-Frye-2dv2} is, of course, quite challenging.  The
quantity $d^3N/dYd\pt^2$ measured in experiment is a result of a (linear)
integral transform in Eq.~(\ref{eq:Cooper-Frye-2dv2}) of functions
$\rho_{1,2}(\alpha)$ characterizing the shape of the freeze-out
hyper-surface and the flow on it.  This information about the
freeze-out surface is blurred and distorted by the integral
transformation.

Furthermore, as the number of points in particle rapidity space
measured by experiments is typically smaller than the number of points
needed in fluid rapidity space to characterize the freeze-out surface,
i.e., the functions $\rho_{1,2}(\alpha)$, or $\tau_f(\a), \eta_f(\a)$,
there would, in principle, be many different freeze-out surfaces which
produce similar phase space particle distributions matching experimental data.

Fortunately, there exists a very well developed method for solving such a
deconvolution problem -- the
maximum entropy method (MEM)\cite{bryan}.  In the spirit of the MEM, if
there are many possible freeze-out surfaces in agreement with data,
then the sensible question one could ask is what is the probability
distribution and the measure in the space of all such freeze-out surfaces
that takes into account the experimental data as well as our prior
expectation of the freeze-out surface. If such probability
distribution is given, the reconstructed freeze-out surface, i.e., the
functions $\tau_f(\a),{\eta}_f(\a)$, can be obtained by averaging over
all possible configurations weighted by the probability density.  
To make our paper self-contained, we provide a brief introduction to MEM in
Appendix~\ref{sec:MEM_review}.

We use the extended version of MEM\cite{bryan,Bajkova:1992,ding} appropriate for the present purpose.
With the aid of MEM,
we reconstruct the longitudinal freeze-out surface,
i.e., we find the ``expectation value" for $\rho_{1,2}$ and thus $\tau_f(\a),{\eta}_f(\a)$ 
given the rapidity-dependent distribution measured by experiment.
Those results are presented in detail in Sec.~\ref{sec:surface} . 
All technical details of maximum entropy reconstruction of freeze-out surface are summarized
in Appendix~\ref{sec:MEM_detail}.

\section{General solutions to $1+1$ ideal hydrodynamics}
\label{sec:hydro_solution}

Once $\tau_f(\a)$ and ${\eta}_f(\a)$
are obtained (by MEM),
our next task is to solve hydrodynamic equations and evolve the system
back in time,
from freeze-out to early times.
It is convenient to change the coordinates in the hydrodynamic equations
from $t$ and $z$ (or $\tau$ and $\eta$) 
to temperature $T$ and fluid rapidity $\alpha$. 
Here, $T$ and $\alpha$ play the role of temporal and spatial variables
respectively (e.g., in the Bjorken flow $T$ is a function of $\tau$ only and
$\alpha=\eta$).
As a result, 
one could recast $1+1$ ideal hydrodynamic equations into one 
\textit{linear} second order differential equation, 
known as \kh~equation~\cite{kh:1954}.
For completeness,
we shall first review the derivation of this equation following Belenkij and Landau\cite{Belenkij:1956cd}.
After that,
we shall study general solutions to that equation with Cauchy initial condition.

\subsection{Khalatnikov equation}
The relativistic ideal hydrodynamic equations
 we study are given by\cite{landau1959fluid}:
\bes
\label{eq:hydro}
\begin{equation}
\label{eq:DT}
\(u^{\mu}\pd_{\mu}\)\epsilon + (\e+p) \(\pd_{\mu} u^{\mu}\) = 0\,   ,
\end{equation}
\begin{equation}
\label{eq:DT2}
(\e+p) \(u^{\nu}\pd_{\nu}\) u^{\mu} 
 + \(g^{\mu\nu} + u^{\mu}u^{\nu} \)\pd_{\nu}p = 0\, ,
\end{equation}
\ees
where $\e, p$ are the energy density and pressure respectively. 
$u^{\mu}$ is the flow four-velocity that obeys $u^{\mu}u_{\mu}=-1.$
Using thermodynamic relations
$d\epsilon = T ds, dp = s dT $,
$\epsilon + p = T s$,
 where $s$ is the entropy density,
Eqs.~\eqref{eq:hydro} can be written as
\bes
\label{eq:hydroA}
\begin{equation}
\pd_{\mu} (s u^{\mu}) = 0\, ,
\end{equation}
which is the conservation of entropy, and
\begin{equation}
u^{\mu}
\[\,\pd_{\mu}(Tu_{\nu}) - \pd_{\nu}(Tu_{\mu})\, \]
= 0 \, .
\end{equation}
\ees
-- the relativistic analog of Euler equation.

For $1+1$ dimensional flow we consider, Eqs.~\eqref{eq:hydroA} become:
\bes
\be
\label{eq:hydro1}
\pd_{t}(\,  s\, u^{t} \,)
+ \pd_{z}(\, s\, u^{z} \,)
= 0 \, ;
\ee
\be
\label{eq:hydro2}
\pd_{t}(\, T u_{z} \,)
- \pd_{z}(\, T u_{t} \,)
= 0\, .
\ee
\ees
Due to \eq\eqref{eq:hydro2},
one can introduce a potential $\psi(t,z)$ such that
\be\label{eq:psi}
d\psi(t,z) = Tu_t\, dt + Tu_z\, dz
= T(-\cosh\a \, dt + \sinh\a\, dz)\, . 
\ee
where we used $\alpha$ defined in Eq.~(\ref{eq:utuz}).
To change the variables from $t,z$ to $T,\a$,
we now introduce a Legendre transform $\psi(t,z)$ of the potential
$\chi(T,\alpha)$ as
\be\label{eq:chi=}
\chi(T, \a) = \psi(t, z) - Tu_t\, t - T u_z\, z
\ee
so that
\begin{multline}
  \label{eq:chi}
  d\chi(T,\a) = -t\, d(Tu_t) -z\, d(T u_z) \\
= \(t\, \cosh\a -
  z\,\sinh\a \)dT \\+ T\(t\, \sinh\a - z\, \cosh\a \)d\a \, .
\end{multline}
The new potential, $\chi(T,\a)$, sometimes referred to as \kh~potential, 
depends on $T,\a$ only.

We then change the variables in \eq\eqref{eq:hydro1} from $t,z$ to
$T,\a$:
   \begin{multline}
   \label{eq:hydroeq2Talpha}
 0=  \frac{\pd(t, z)}{\pd (T, \a)}\,
   \[\,\frac{\pd(s \,\cosh\a,z)}{\pd (t,z)}+\frac{\pd(s\, \sinh\a, t)}{\pd(z,t)}\,\]
  \\ = \frac{\pd(s \,\cosh\a,z)}{\pd( T,\a)}
   -\frac{\pd(s\, \sinh\a, t)}{\pd(T,\a)}\\
   =\frac{ds}{dT}\[\,-\frac{\pd t}{\pd \a}\sinh\a+\frac{\pd z}{\pd \a}\cosh\a\,\]\\
- s\[\,-\frac{\pd t}{\pd T}\cosh\a+\frac{\pd z}{\pd T}\sinh\a\,\]
 \, ,
   \end{multline}
   where $\pd(t,z)/\pd (T,\a)$ denotes the Jacobian of the variable
   transformation from $(t,z)$ to $(T,\a)$.  Using \eq\eqref{eq:chi}
   to simplify \eq\eqref{eq:hydroeq2Talpha}, we arrive at a second
   order linear partial differential equation for
   $\chi(T,\a)$~\cite{kh:1954,Belenkij:1956cd}\footnote{Equation.~\eqref{eq:kheq} has been applied to study longitudinal expansion in heavy-ion collisions 
in Ref.~\cite{Bialas:2007iu,*Beuf:2008vd,*Peschanski:2010cs} recently.}
   \begin{equation}
\label{eq:kheq}
\left[c_s^2T^2\pd_T^2 + T\pd_T - \pd_\a^2\right]\chi(T,\a)=0,
\end{equation}
where $c_s$ is the ($T$-dependent) speed of sound:
\begin{equation}
  \label{eq:cs}
  c_s^2 = \frac{dp}{d\epsilon}= \frac {sdT}{Tds}.
\end{equation}

Once the potential $\chi(T,\alpha)$ is found using Khalatnikov
equation~(\ref{eq:kheq})
one can determine  $\tau(T,\a)$ and ${\eta}(T,\a)$ from the
derivatives of $\chi$:
\bes
\label{eq:chitauomega}
\begin{equation}
\label{eq:chicosh}
\pd_{T}\chi(T, \a)
=  \tau \cosh(\alpha-\eta)
=  \tau  u^{\tau} \, ;
\end{equation}
\begin{equation}
\label{eq:chisinh}
  \pd_{\a}\chi(T, \a)
=
 T\tau\sinh(\alpha-\eta)
  =   T\tau^{2}  u^{\eta}\, , 
\end{equation}
\ees
where we used Eqs.~\eqref{eq:chi},~(\ref{eq:taueta}) and~(\ref{eq:utaueta}).

To see how Khalatnikov equation~\eqref{eq:kheq} works,
it is instructive to check it against known Bjorken solution in which 
$\a=\eta$ or $u^{\eta} =0$.
On that solution, $\chi$ is independent of $\a$ 
according to \eq\eqref{eq:chisinh}.
Substituting $\chi(T)$ into \eq\eqref{eq:kheq},
one finds, for constant $c_s$,
$\chi(T)= C_1+C_2 T^{1-c_s^{-2}}$ where $C_1,C_2$ are integration constants.
Further substituting this into \eq\eqref{eq:chicosh},
we have $\tau\sim T^{-1/c^2_s}$ --
the well-known result of Bjorken \cite{Bjorken:1982qr}.   

 \subsection{The general solution of Cauchy problem for Khalatnikov equation}
 \label{sec:generalsol}
 
 We now turn to the general solution of \eq\eqref{eq:kheq} with Cauchy
 initial condition on an isothermal hypersurface $T(t,z)=\bar T$.  
For simplicity, we shall consider the case of constant~$c_s$. It is convenient to introduce a new variable 
 \begin{equation}
y =
 \log (T/\bar T)\,. \label{eq:yT}
\end{equation}
It is also convenient to introduce a rescaled potential
 $\bar\chi(y;\a)$, related to $\chi(T;\a)$ by
 \begin{equation} \chi(T; \a) =
 e^{-\nu y}\bar\chi(y; \a)\,, 
\qquad\mbox{where}\quad
\n = \frac{1}{2}\left(c^{-2}_s- 1\right)\, .
   \end{equation}
In terms of $\bar\chi(y;\a)$, the
 \kh~equation takes the form of a massive Klein-Gordon equation:
 \be
   \label{eq:eqkh1}
   \(\,\pd^2_{y} -c^{-2}_{s}\pd^2_\a -\nu^2\,\)\bar\chi(y;\a) =0\, .
   \ee

   To solve \eq\eqref{eq:eqkh1},
   we introduce a Green's function $G(y;\a)$
  {in terms of the modified Bessel function $I_0$},
   \begin{multline}
     \label{eq:green-kh}
     G(y;\a) =  \frac{c_{s}}{2}\, I_{0}(\nu\sqrt{y^{2}-c^2_s\a^{2}})
\\\times
     \[\,\theta(y-c_{s}\a)-\theta(-y-c_{s}\a)\,\]\, ,
   \end{multline}
   which satisfies \eq\eqref{eq:eqkh1}
   with the Cauchy initial condition
   \begin{equation}
     \label{eq:boundary-g}
     \lim_{y\to 0} G(y;\a) = 0\, ,
     \qquad
     \lim_{y\to 0}\frac{\pd G(y;\a)}{\pd y} = \delta({ \alpha})\, . 
   \end{equation}
We note that
$G(y;\a)$ vanishes outside of the ``sound horizon" at $\a=\pm y/c_s$.
Solutions with given initial values of $\bar\chi(y,\alpha)$ and
$\pd_y\bar\chi(y,\alpha)$
on the iso-thermal surface $y=0$ can then be found using $G(y;\a-\a')$:
\begin{equation}
\label{eq:chisol}
        \bar\chi(y,\a)
=  \int^{\infty}_{-\infty}\!\!d\a'
\[\,
\bar\chi(0,\a')\pd_{y} + \pd_{y}\bar\chi(0,\a')\,\]
G(y;\a-\a' )\, .
\end{equation}

\section{Results}
\label{sec:results}

\subsection{Reconstructed freeze-out surface}
\label{sec:surface}

 \begin{figure}[htb]
  \centering
	\includegraphics[width=.45\textwidth]{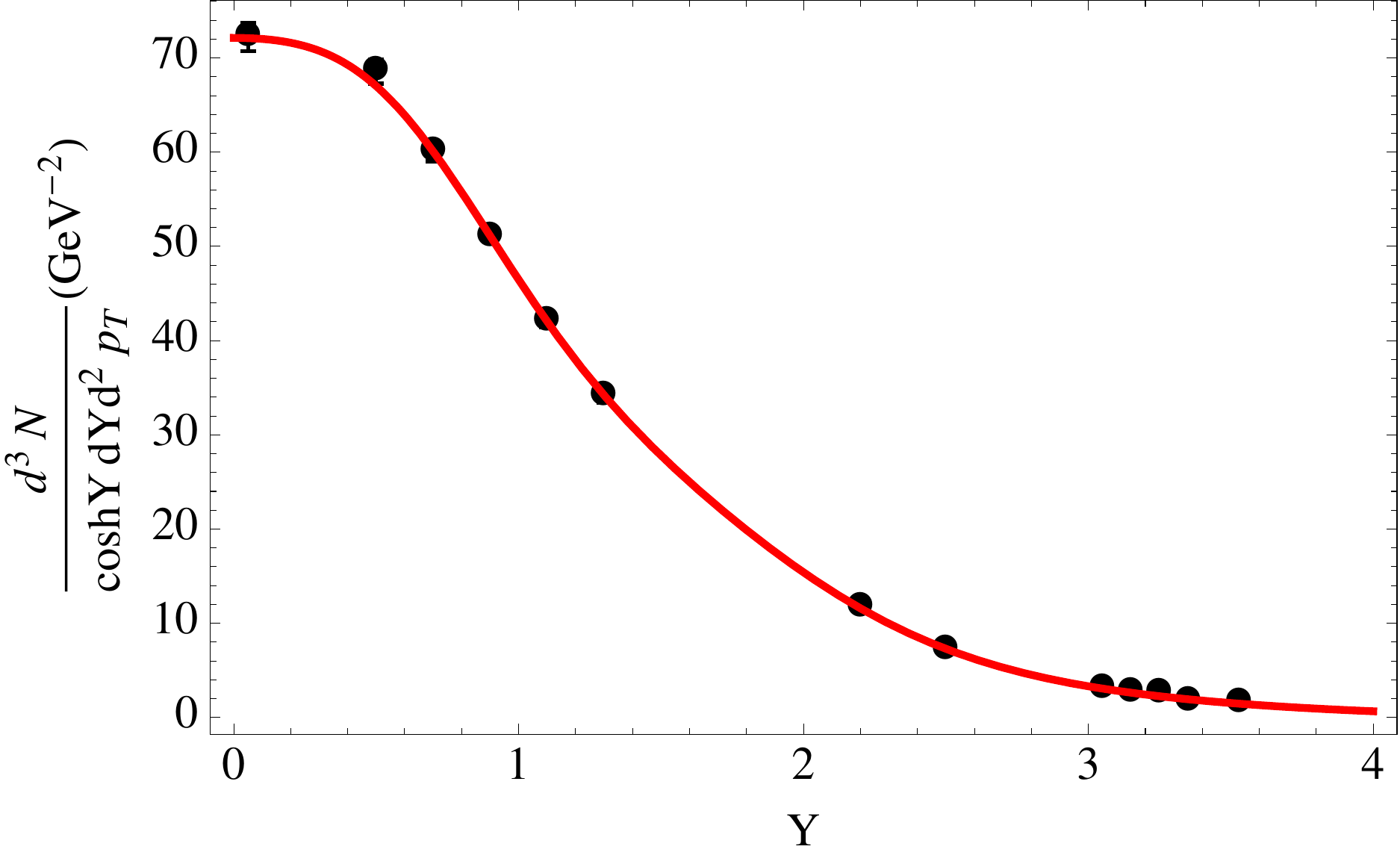}
    \caption{
  	    (Color online)  	    
Charged pion rapidity distribution 
in central Au-Au collisions at $\sqrt{s}=200$GeV in the $\pt$ bin at
$\pt=0.55$GeV~\cite{Bearden:2004yx}. 
The experimental data errors are similar to the size of data points.
{The red curve is the rapidity distribution obtained from the
  MEM-reconstructed freeze-out surface (see
  Fig.~\ref{fig:freeze_out_surface} and text).}
       }
\label{fig:dNdY}
\end{figure}

\begin{figure}	
	\includegraphics[width=.45\textwidth]{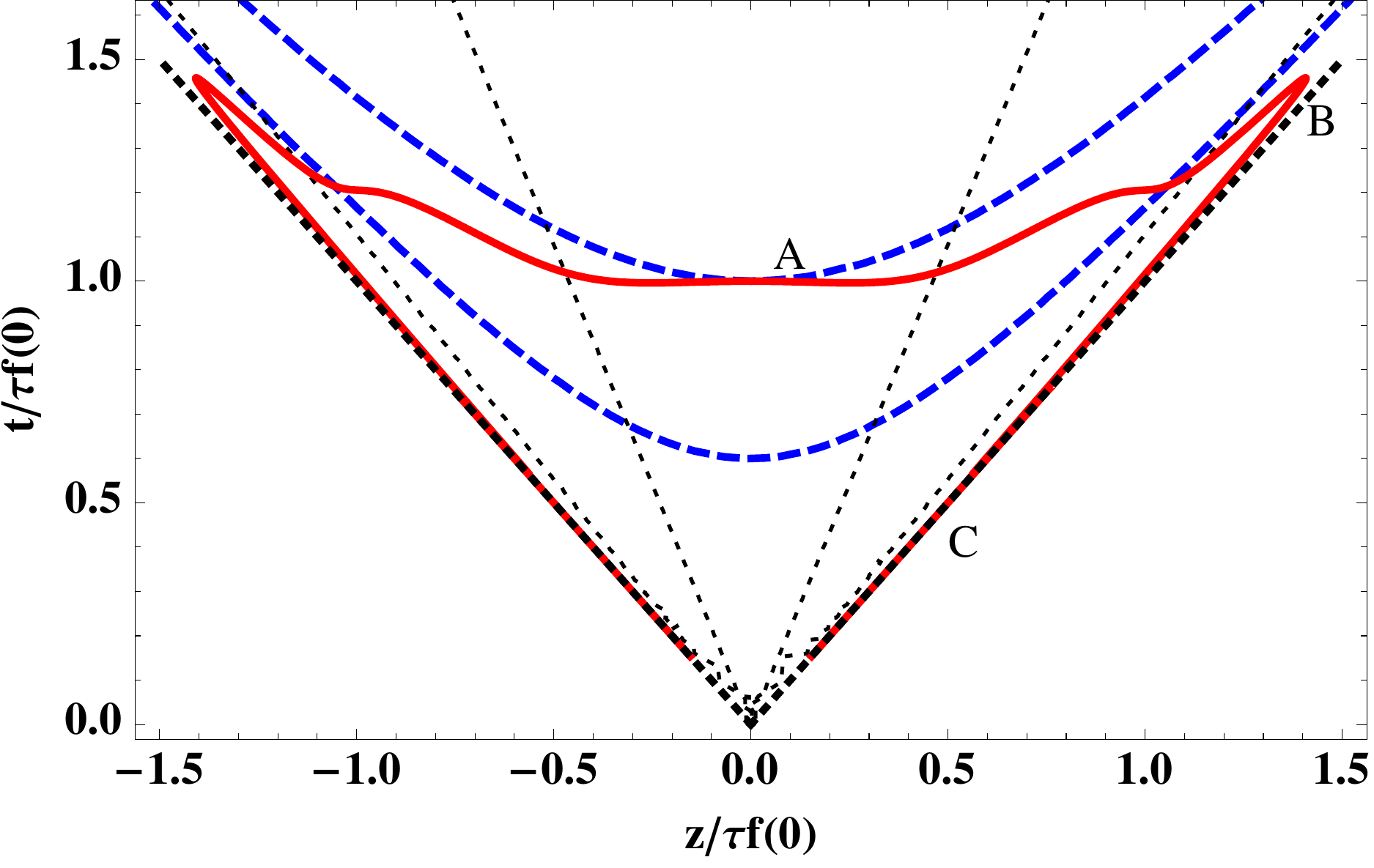}
\caption{
 (Color online) MEM-reconstructed freeze-out surface (red curve). 
Bjorken proper time $\tau={\rm const}$ curves are dashed blue.   
Dotted rays correspond to Bjorken rapidities $\eta=\pm1,\pm2$.
Letters A, B and C label characteristic points on the freezeout surface
(see text {and Fig.~\ref{fig:tauetaf}}). The reconstruction
uncertainties (not shown) are discussed in Appendix~\ref{sec:MEM_detail}.
}
\label{fig:freeze_out_surface}
\end{figure}

We first present our results on the MEM reconstruction of the
longitudinal freeze-out surface and flow.  Assuming the freeze-out
temperature $\Tf=140$~MeV, which is close to the one used in
hydrodynamic simulations (e.g. \cite{Schenke:2010nt,Hirano:2001yi}),
we have applied the maximum entropy method to the pion
rapidity distribution measured in Au-Au central collisions
at $\sqrt s=200$ GeV~\cite{Bearden:2004yx}. For this analysis we chose
particles within the transverse momentum bin of width 
{0.1~GeV} centered at
$p_\perp=0.55$~GeV. 
This choice is motivated by the desire to minimize
the effect of resonances (at lower $p_\perp$) and
viscous or non-hydrodynamic corrections (at higher~$p_\perp$).

The experimental data are plotted using black dots
in Fig.~\ref{fig:dNdY}.  
Applying the MEM to these data points, 
we obtain the reconstructed freeze-out surface shown in Fig.~\ref{fig:freeze_out_surface}. 
As an important check, 
we also input the reconstructed freeze-out surface back into  the Cooper-Frye formula \eqref{eq:Cooper-Frye-2dv2} 
and compute the corresponding rapidity distribution. 
The result is shown in Fig.~\ref{fig:dNdY},
indicating that our reconstructed freeze-out surface is consistent with the data.

In Fig.~\ref{fig:freeze_out_surface}, we show the reconstructed freeze-out
surface in $t-z$ coordinates using the red solid curve.  We rescaled
dimensionful quantities such as $\tf(\a), \zf(\a)$ by $\tau_f(0)$ --
the proper time at which the center of the fireball freezes out.  As
we have not included the dynamics of radial expansion in our analysis,
we could only estimate the combination $A_{\perp}\tau_f(0)$ rather than
 $A_{\perp}$ and $\tau_f(0)$ individually. We find
$A_{\perp}\tau_f(0)\approx (1.4-1.5)\times 10^4\ {\rm fm}^3$ (somewhat large compared to typical values in 
hydrodynamic simulations).

\begin{figure}[htb]
  \centering
	\includegraphics[width=25em]{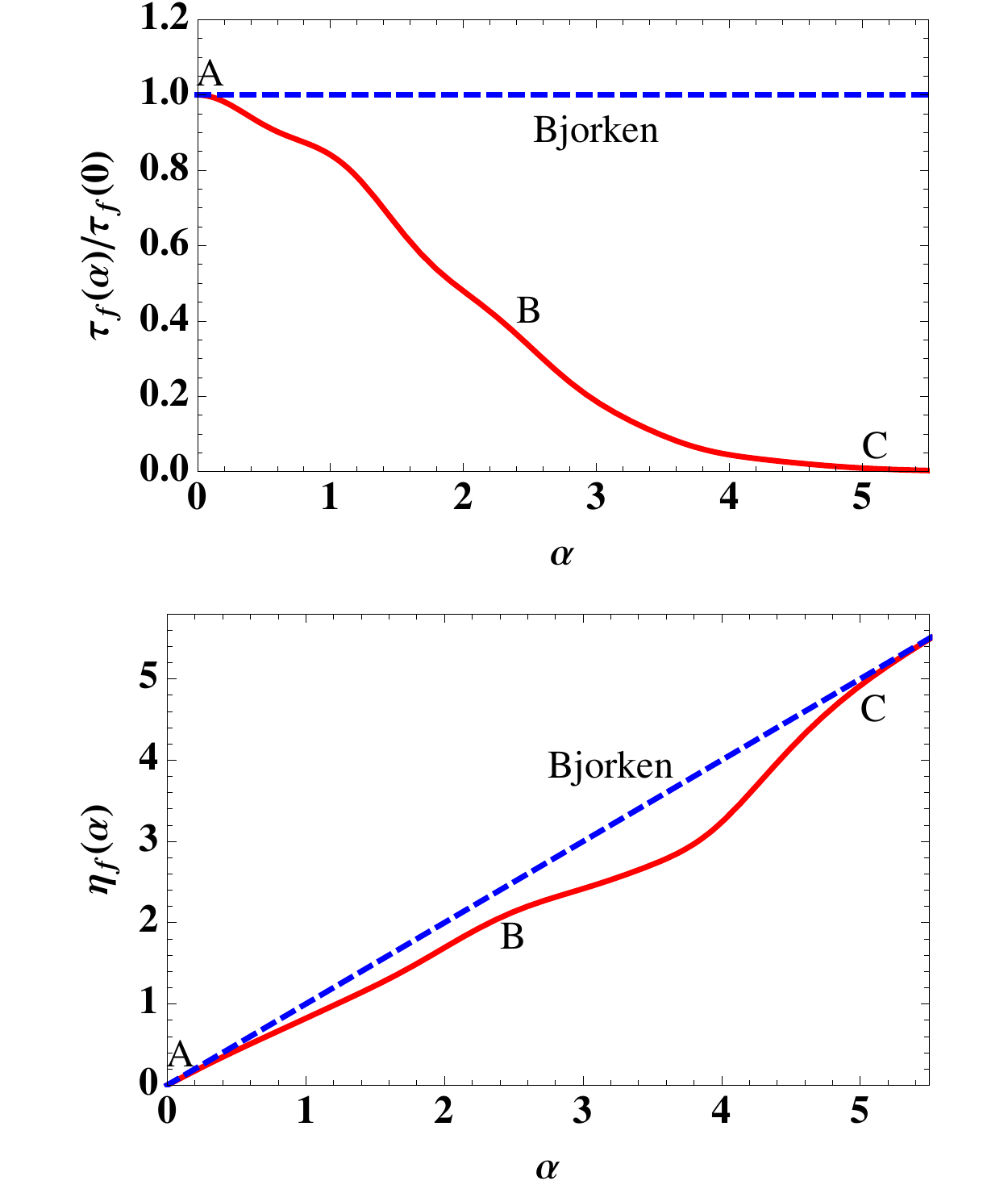}
\caption{
(Color online) The shape, $\tau_f(\a)/\tau_f(0)$, and the flow
profile, $\eta_f(\a)$, of the freeze-out surface found
using MEM, as in Fig.~\ref{fig:freeze_out_surface}, but represented
parametrically in Bjorken
coordinates. Letters A, B and C label
characteristic points on the freeze-out surface and correspond to
those in Fig.~\ref{fig:freeze_out_surface}.  Dashed blue lines show
Bjorken's boost-invariant approximation.  }
\label{fig:tauetaf}
\end{figure}

The difference between our reconstructed freeze-out surface 
and idealized Bjorken's boost-invariant model, where freeze-out occurs
on an equal-proper-time hyperbola, is noticeable 
in Fig.~\ref{fig:freeze_out_surface}.
This difference is more clearly illustrated in Fig.~\ref{fig:tauetaf}
where freeze-out proper time $\tau_f(\a)$ as a function of fluid
rapidity $\a$ is plotted as red solid curve. This plot shows that the
mid-rapidity region of the fireball freezes out at later Bjorken proper
times than the forward/backward rapidity regions.

Fig.~\ref{fig:tauetaf} shows the longitudinal flow profile on the
freeze-out surface. Since $\alpha>\eta$, the flow rapidity is faster
than the idealized boost-invariant Bjorken scenario. This is expected,
given the pressure gradients due to the pressure decreasing away from
mid-rapidity.
The effect in terms of
$|\(\a-\eta_{f}\)/\eta_{f}|$ is of the order of $10\% - 20\%$ for all $\a$ we are considering.

Figure \ref{fig:tauetaf} could be thought of as a parametric
representation of the freeze-out curve in $t-z$ coordinates shown in
Fig.~\ref{fig:freeze_out_surface}. 
As the parameter $\alpha$ increases
from 0, the corresponding point in $tz$ plane traces the curve from A
to B to C. These points are also marked in Figure~\ref{fig:tauetaf}. For example, one can see that the increase of $z$
from A to B and subsequent decrease from B to C is due to the
competition between falling $\tau(\alpha)$ and rising $\eta(\alpha)$
in the formula $z=\tau\sinh\eta$.

We have, therefore, reconstructed hydrodynamic conditions on the
freeze-out surface directly from the experimental data, remarkably, without
using hydrodynamic simulations.  We shall now take this result as the
``initial'' condition to evolve the system back in time.

\subsection{Temperature and flow history}
\label{sec:longitudinalevolution}

\begin{figure}[htb]
  \centering
	\includegraphics[height=20em]{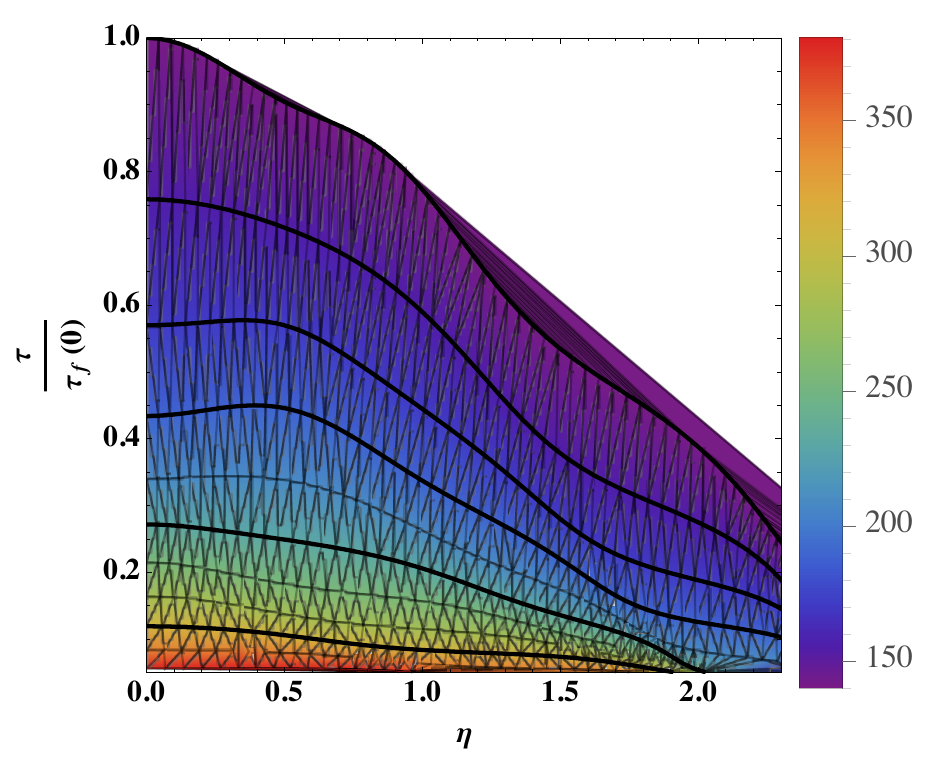}
%
\caption{
(Color online) The temperature {(in~MeV)}
map in Bjorken coordinates.
Solid contours denote iso-thermal surfaces.
}
\label{fig:evolution}
\end{figure} 

We now present our results on the longitudinal evolution of the fireball.
They are obtained by using the analytic solution to $1+1$ ideal
hydrodynamic equation, with Cauchy initial condition
namely, \eq\eqref{eq:chisol}.
To use our knowledge on freeze-out surface,
i.e., $\tau_f(\a),\, {\eta}_f(\a)$, 
we set $\bar T$ to be $T_f$ in our definition of $y$, i.e.,
$y=\log(T/T_f)$ in \eq\eqref{eq:chisol}. The fact that the freeze-out
surface is at $T={\rm const}$ makes this boundary condition easy to set.
As a result,
we have from \eq\eqref{eq:chitauomega}
\bes\label{eq:chi1coshsinh}
\begin{equation}
\label{eq:chi1cosh}
\[-\nu \bar\chi(0;\a) + \pd_y\bar\chi(0;\a)\]
= T_f\tau_f(\a)\cosh(\alpha-\eta_f(\alpha));
\end{equation}
\begin{equation}
\label{eq:chi1sinh}
\pd_\a\bar\chi(0;\a)=
T_f\tau_f(\a)\sinh(\alpha-\eta_f(\alpha))
{\, .}
\end{equation}
\ees We then determine $\bar\chi(0;\a)$ by integrating
\eq\eqref{eq:chi1sinh} using $\tau_f(\a)$ and ${\eta}_f(\a)$ obtained
by MEM reconstruction and shown in Fig.~\ref{fig:tauetaf}. In fact, as
seen in Eqs.~(\ref{eq:rho12_def}), the MEM directly outputs functions appearing in
the r.h.s. of Eqs.~(\ref{eq:chi1cosh}) and~(\ref{eq:chi1sinh}).  The
integration constant is unphysical since {  
  $\bar\chi(0,\alpha)=\chi(T_f,\alpha)$, and potential $\chi$ is defined up to
  a constant, Eqs.~\eqref{eq:psi},~\eqref{eq:chi=}. It
can be fixed by requiring $\bar\chi(0;\a)\to0$ when
$\alpha\to\pm\infty$.  We next substitute the resulting
$\bar\chi(0;\a)$ into \eq\eqref{eq:chi1cosh} to obtain
$\pd_y\bar\chi(0;\a)$.  Consequently, $\bar\chi(y;\a)$ for $y>0$ can
be obtained from explicit solution \eq\eqref{eq:chisol}.  Finally, we
read $\tau(T,\a),\eta(T,\a)$ from resulting $\chi=e^{-\nu y} \bar\chi$
using \eq\eqref{eq:chitauomega}.

The longitudinal expansion history we find is summarized in Fig.~\ref{fig:evolution}.
We used conformal equation of state $c^2_s=1/3$.
We also checked that the output with a different choice of $c^2_s$ is qualitatively similar.

Iso-thermal surfaces are shown as black solid curves in
Fig.~\ref{fig:evolution}. For comparison, Bjorken solution would
correspond to horizontal isothermal contours on such a plot. We note
that, even though the freeze-out {(isothermal)} surface has no visible
plateau, the isothermal surfaces at higher temperatures (earlier
times) do display a plateau which shrinks as the system evolves (cools down).
We will comment more on this behavior later using different graphic 
representations.

It is interesting to compare the temperature (energy density) and flow
profile at early (proper) time with initial conditions used in
hydrodynamic
simulations~\cite{Hirano:2001yi,Satarov:2006iw,Bozek:2009ty,Schenke:2010nt}.
In such
simulations
initial energy profile is typically assumed to be flat around
mid-rapidity with half a Gaussian fall-off in the forward and backward
rapidity directions.
Moreover, $v_{\eta}$ is set to zero initially.

\begin{figure}[htb]
  \centering	
	\includegraphics[width=25em]{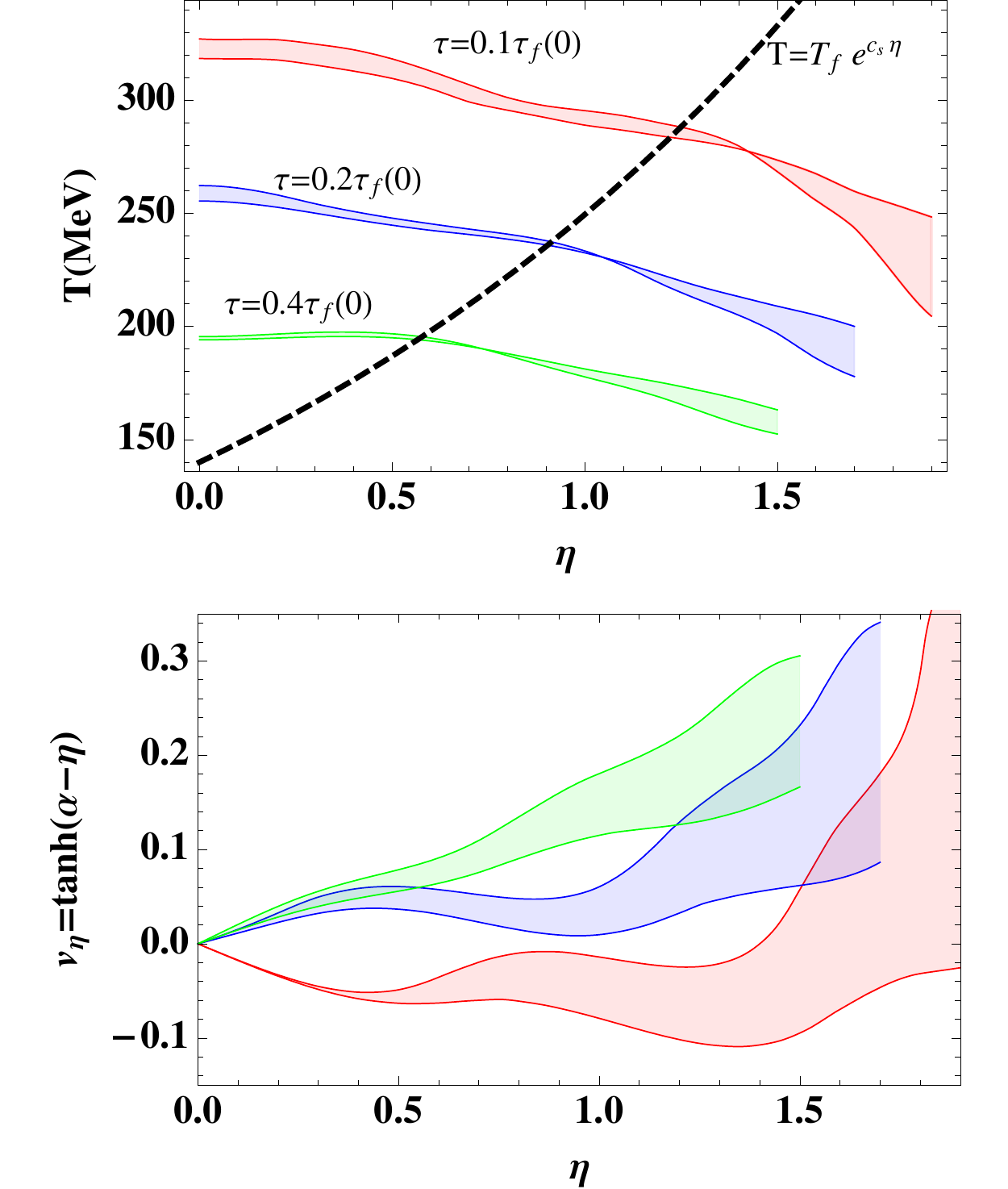}	
\caption{ 
(Color online)
 Temperature and flow profile on various equal-proper-time surfaces.
Red, blue and green curves correspond to $\tau=0.1, 0.2, 0.4\tau_{f}(0)$ respectively.
Shaded bands illustrate the initial condition dependence due to
  sensitivity to default model choice in MEM 
(see Appendix.~\ref{sec:MEM_detail}).}
\label{fig:equaltau}
\end{figure} 

To facilitate the comparison, in Fig.~\ref{fig:equaltau},
we plot equal proper time temperature profile, $T$ vs $\eta$,
and flow profile, $v_{\eta}$ vs $\eta$,
at early times: $\tau=0.4\tau_f(0)$, $0.2\tau_f(0)$, and
$0.1\tau_f(0)$.  In accordance with our observations in
Fig.~\ref{fig:evolution}, the early-time temperature profile in
Fig.~\ref{fig:equaltau} is qualitatively similar to the Bjorken-like
plateau truncated at forward and backward rapidities used in
hydrodynamic
simulations~\cite{Hirano:2001yi,Satarov:2006iw,Bozek:2009ty,Schenke:2010nt}.
In Fig.~\ref{fig:equaltau} one can also see that the flow is almost
Bjorken-like ($v_\eta\equiv\tanh(\alpha-\eta)\approx0$) within the
temperature plateau.  In particular, the early-time profile we observe
is much closer to Bjorken boost-invariant model ($\alpha=\eta$), than
to the full stopping assumption ($\alpha=0$).

Interestingly, at the earliest times we considered, $v_\eta$ appears
to be slightly negative (i.e., $\alpha<\eta$) within the plateau.
Given the size of the uncertainties in the reconstruction of the flow
(see bands on Fig.~\ref{fig:equaltau} and discussion below) we must
interpret this result cautiously.  It would be interesting to
investigate this question further. Such a slower relative to Bjorken
(Hubble-like) flow would be a natural consequence of the negative
longitudinal pressure in the pre-equilibrium glasma stage of heavy-ion
collisions~\cite{Lappi:2006fp}. Early-time negative longitudinal
pressure also occurs in holographic models of colliding
shock waves~\cite{Grumiller:2008va}, and has been observed in the
``complex deformation'' of the Bjorken flow studied in
Ref.~\cite{Gubser:2012gy}, where $v_\eta<0$ ($\alpha<\eta$) can be
also seen in its Fig.~2.

It is easy to see that the shrinking of the early-time Bjorken plateau
follows the inward propagation of the shock waves from the plateau's
boundaries. Since the local velocity of the wave is $c_s$, it
is easy to find that the wave propagates on top of the Bjorken flow in
Bjorken coordinates according to $d\eta/d\ln\tau=c_s$.  The length of
the plateau shrinks linearly in $\ln\tau$ at a rate of (twice) of
$c_s$.  Since, under Bjorken flow, $T\sim\tau^{-c_s^2}$, the
edge of the plateau should follow $T\sim e^{c_s\eta}$ in the 
{$T-\eta$}
plot as illustrated by the dashed
  black curve in Fig.~\ref{fig:equaltau}.

To estimate the sensitivity of our results to the uncertainty of the
  reconstruction of the freeze-out surface (``initial'' conditions), 
we have reconstructed freeze-out surface using different choices of
default models in MEM {(see Appendix  \ref{sec:MEM_detail})} and evolved
the output of such a reconstruction back in time.
Shaded bands in Fig.~\ref{fig:equaltau} illustrate the difference due
to the variation of default model.
We observe that the temperature (energy density) profile due to the MEM
reconstruction and back-in-time evolution is quite robust against the
change of the default model.
However, the uncertainty in the flow profile is  larger.
Some more details on the default model dependence are given in Appendix.~\ref{sec:MEM_detail}.

\section{Conclusion}
\label{sec:conclusion}

In this paper, we studied longitudinal expansion in heavy-ion
collisions beyond Bjorken boost-invariant model.  In contrast to
existing body of work on the subject, we considered hydrodynamic
evolution back in time -- from freeze-out to earlier times. This is
possible because we use experimental data directly to determine
the starting point of the back-in-time evolution. To achieve that we
applied the maximum entropy method to invert the Cooper-Frye freeze-out
integral transformation relating the shape of the freeze-out
hypersurface surface and the flow on it to rapidity distribution of
observed particles. We use Khalatnikov-Landau approach to solve the
longitudinal hydrodynamic Cauchy problem analytically.

We want to point out two remarkable facts, which together make our
two-step approach natural and straightforward to implement. First, the
transformation from the pair of functions
$\tau_f\cosh(\alpha-\eta_f)=\rho_1/A_\perp$ and
$\tau_f\sinh(\a-\eta_f)=\rho_2/A_\perp$ to experimentally observable
$d^3N/dYd^2p_\perp$ given by Eq.~(\ref{eq:Cooper-Frye-2dv2}) is {\em
  linear}. Second, the same pair of functions translate linearly into
the initial conditions, Eqs.~(\ref{eq:chi1coshsinh}), for the
Khalatnikov equation which, in turn, is also linear.

By reconstructing the freeze-out surface and evolving the system back
in time, we obtained both the temperature and the flow profile history
of the longitudinal expansion.  
The temperature profile features a
Bjorken-like plateau at early times. The plateau shrinks as the
boundary shocks propagate inward, towards the center of the fireball
(see Fig.~\ref{fig:equaltau}). 
The flow velocity $v_\eta$ with respect to local Bjorken frame remains
small within the shrinking plateau (Fig.~\ref{fig:equaltau}), i.e.,
the flow is close to being Bjorken-like (Hubble-like) inside the
plateau: flow rapidity is close to Bjorken rapidity $\alpha\approx
\eta$. The flow rapidity increases faster than Bjorken rapidity
outside of the plateau  ($\alpha>\eta$ or
$v_\eta>0$) due to pressure gradients. 

We also
observe the evidence that $v_\eta<0$ within the plateau at earlier
times, which could represent the effect of the negative longitudinal pressure
in pre-equilibrium glasma stage.
These results provide useful information for determining the initial
conditions for fully numerical hydrodynamic simulations as well as
 holographic models of heavy-ion collisions~\cite{Chesler:2010bi,Casalderrey-Solana:2013aba}.

In this first exploratory study of the new back-in-time approach
to fireball evolution, we attempted to achieve the most analytically
transparent, yet phenomenologically meaningful results.  For this
reason, we focused on longitudinally expansion and neglected many other
potentially important effects. Most notably, we neglected the dynamics
of the transverse expansion and assumed temperature independent sound
speed. Although not realistic, this is a common and useful
approximation, successfully used since the seminal papers by Landau and
Bjorken. 
Relaxing these approximations is, however, straightforward
and extending the present approach to study the hydrodynamic expansion
beyond longitudinal expansion would be an interesting direction to pursue.

We also neglected the effects of viscosity. Given the almost perfect
fluidity of the quark-gluon plasma in the regime probed by heavy-ion
collisions,
the effect of viscosity should be small as far as
  bulk hydrodynamics we study is concerned. We believe that the
advantage of analytical transparency afforded by the ideal
hydrodynamic approximation more than compensates for the lack of
numerical accuracy. Our goal is to provide useful insights for fully
numerical hydrodynamic approaches aiming at quantitative precision.

One of the new ingredients in our analysis is the application of
maximum entropy method (MEM).  We found it to be a useful tool for
inverting the Cooper-Frye freeze-out and reconstructing the freeze-out
surface.  As a general deconvolution method, MEM may also be used
to reconstruct freeze-out surface in transverse plane or even full
$3$d freeze-out surface.  The result of such a 2d or 3d reconstruction might
reveal interesting information about the structure and evolution of
the heavy-ion fireball.
   
\acknowledgments

The authors would like to express their gratitude to Paul Chesler,
Ulrich Heinz, Dmitri Kharzeev, Denes Molnar, Wilke van der Schee,
Edward Shuryak, Dam Son, Derek Teaney, Raju Venugopalan and Ho-Ung Yee
for useful comments and discussions.  Y.Y. would like to especially
thank Todd Springer for stimulating conversations, Heng-Tong Ding for
discussing MEM reconstruction, Kolja Kauder for explaining features of
experimental measurements, and to acknowledge the
lessons learned from the study of the lattice QCD MEM C-codes written
by Pavel Buividovich.  Y.Y. is grateful to UIC Dean's Scholar
fellowship program for support.  This research is supported by the US
DOE grant No. DE-FG0201ER41195.

\bibliography{MEM_hydro}

\begin{thebibliography}{30}%
\makeatletter
\providecommand \@ifxundefined [1]{%
 \@ifx{#1\undefined}
}%
\providecommand \@ifnum [1]{%
 \ifnum #1\expandafter \@firstoftwo
 \else \expandafter \@secondoftwo
 \fi
}%
\providecommand \@ifx [1]{%
 \ifx #1\expandafter \@firstoftwo
 \else \expandafter \@secondoftwo
 \fi
}%
\providecommand \natexlab [1]{#1}%
\providecommand \enquote  [1]{``#1''}%
\providecommand \bibnamefont  [1]{#1}%
\providecommand \bibfnamefont [1]{#1}%
\providecommand \citenamefont [1]{#1}%
\providecommand \href@noop [0]{\@secondoftwo}%
\providecommand \href [0]{\begingroup \@sanitize@url \@href}%
\providecommand \@href[1]{\@@startlink{#1}\@@href}%
\providecommand \@@href[1]{\endgroup#1\@@endlink}%
\providecommand \@sanitize@url [0]{\catcode `\\12\catcode `\$12\catcode
  `\&12\catcode `\#12\catcode `\^12\catcode `\_12\catcode `\%12\relax}%
\providecommand \@@startlink[1]{}%
\providecommand \@@endlink[0]{}%
\providecommand \url  [0]{\begingroup\@sanitize@url \@url }%
\providecommand \@url [1]{\endgroup\@href {#1}{\urlprefix }}%
\providecommand \urlprefix  [0]{URL }%
\providecommand \Eprint [0]{\href }%
\providecommand \doibase [0]{http://dx.doi.org/}%
\providecommand \selectlanguage [0]{\@gobble}%
\providecommand \bibinfo  [0]{\@secondoftwo}%
\providecommand \bibfield  [0]{\@secondoftwo}%
\providecommand \translation [1]{[#1]}%
\providecommand \BibitemOpen [0]{}%
\providecommand \bibitemStop [0]{}%
\providecommand \bibitemNoStop [0]{.\EOS\space}%
\providecommand \EOS [0]{\spacefactor3000\relax}%
\providecommand \BibitemShut  [1]{\csname bibitem#1\endcsname}%
\let\auto@bib@innerbib\@empty
\bibitem [{\citenamefont {Bjorken}(1983)}]{Bjorken:1982qr}%
  \BibitemOpen
  \bibfield  {author} {\bibinfo {author} {\bibfnamefont {J.}~\bibnamefont
  {Bjorken}},\ }\href {\doibase 10.1103/PhysRevD.27.140} {\bibfield  {journal}
  {\bibinfo  {journal} {Phys.Rev.}\ }\textbf {\bibinfo {volume} {D27}},\
  \bibinfo {pages} {140} (\bibinfo {year} {1983})}\BibitemShut {NoStop}%
\bibitem [{\citenamefont {Bearden}\ \emph {et~al.}(2005)\citenamefont {Bearden}
  \emph {et~al.}}]{Bearden:2004yx}%
  \BibitemOpen
  \bibfield  {author} {\bibinfo {author} {\bibfnamefont {I.}~\bibnamefont
  {Bearden}} \emph {et~al.} (\bibinfo {collaboration} {BRAHMS Collaboration}),\
  }\href {\doibase 10.1103/PhysRevLett.94.162301} {\bibfield  {journal}
  {\bibinfo  {journal} {Phys.Rev.Lett.}\ }\textbf {\bibinfo {volume} {94}},\
  \bibinfo {pages} {162301} (\bibinfo {year} {2005})},\ \Eprint
  {http://arxiv.org/abs/nucl-ex/0403050} {arXiv:nucl-ex/0403050 [nucl-ex]}
  \BibitemShut {NoStop}%
\bibitem [{\citenamefont {Back}\ \emph {et~al.}(2005)\citenamefont {Back},
  \citenamefont {Baker}, \citenamefont {Ballintijn}, \citenamefont {Barton},
  \citenamefont {Becker} \emph {et~al.}}]{Back:2004je}%
  \BibitemOpen
  \bibfield  {author} {\bibinfo {author} {\bibfnamefont {B.}~\bibnamefont
  {Back}}, \bibinfo {author} {\bibfnamefont {M.}~\bibnamefont {Baker}},
  \bibinfo {author} {\bibfnamefont {M.}~\bibnamefont {Ballintijn}}, \bibinfo
  {author} {\bibfnamefont {D.}~\bibnamefont {Barton}}, \bibinfo {author}
  {\bibfnamefont {B.}~\bibnamefont {Becker}},  \emph {et~al.},\ }\href
  {\doibase 10.1016/j.nuclphysa.2005.03.084} {\bibfield  {journal} {\bibinfo
  {journal} {Nucl.Phys.}\ }\textbf {\bibinfo {volume} {A757}},\ \bibinfo
  {pages} {28} (\bibinfo {year} {2005})},\ \Eprint
  {http://arxiv.org/abs/nucl-ex/0410022} {arXiv:nucl-ex/0410022 [nucl-ex]}
  \BibitemShut {NoStop}%
\bibitem [{\citenamefont {Abbas}\ \emph {et~al.}(2013)\citenamefont {Abbas}
  \emph {et~al.}}]{Abbas:2013bpa}%
  \BibitemOpen
  \bibfield  {author} {\bibinfo {author} {\bibfnamefont {E.}~\bibnamefont
  {Abbas}} \emph {et~al.} (\bibinfo {collaboration} {ALICE Collaboration}),\
  }\href@noop {} {\  (\bibinfo {year} {2013})},\ \Eprint
  {http://arxiv.org/abs/1304.0347} {arXiv:1304.0347 [nucl-ex]} \BibitemShut
  {NoStop}%
\bibitem [{\citenamefont {Satarov}\ \emph {et~al.}(2007)\citenamefont
  {Satarov}, \citenamefont {Merdeev}, \citenamefont {Mishustin},\ and\
  \citenamefont {Stoecker}}]{Satarov:2006iw}%
  \BibitemOpen
  \bibfield  {author} {\bibinfo {author} {\bibfnamefont {L.}~\bibnamefont
  {Satarov}}, \bibinfo {author} {\bibfnamefont {A.}~\bibnamefont {Merdeev}},
  \bibinfo {author} {\bibfnamefont {I.}~\bibnamefont {Mishustin}}, \ and\
  \bibinfo {author} {\bibfnamefont {H.}~\bibnamefont {Stoecker}},\ }\href
  {\doibase 10.1103/PhysRevC.75.024903} {\bibfield  {journal} {\bibinfo
  {journal} {Phys.Rev.}\ }\textbf {\bibinfo {volume} {C75}},\ \bibinfo {pages}
  {024903} (\bibinfo {year} {2007})},\ \Eprint
  {http://arxiv.org/abs/hep-ph/0606074} {arXiv:hep-ph/0606074 [hep-ph]}
  \BibitemShut {NoStop}%
\bibitem [{\citenamefont {Bozek}\ and\ \citenamefont
  {Wyskiel}(2009)}]{Bozek:2009ty}%
  \BibitemOpen
  \bibfield  {author} {\bibinfo {author} {\bibfnamefont {P.}~\bibnamefont
  {Bozek}}\ and\ \bibinfo {author} {\bibfnamefont {I.}~\bibnamefont
  {Wyskiel}},\ }\href {\doibase 10.1103/PhysRevC.79.044916} {\bibfield
  {journal} {\bibinfo  {journal} {Phys.Rev.}\ }\textbf {\bibinfo {volume}
  {C79}},\ \bibinfo {pages} {044916} (\bibinfo {year} {2009})},\ \Eprint
  {http://arxiv.org/abs/0902.4121} {arXiv:0902.4121 [nucl-th]} \BibitemShut
  {NoStop}%
\bibitem [{\citenamefont {Hirano}\ \emph {et~al.}(2002)\citenamefont {Hirano},
  \citenamefont {Morita}, \citenamefont {Muroya},\ and\ \citenamefont
  {Nonaka}}]{Hirano:2001yi}%
  \BibitemOpen
  \bibfield  {author} {\bibinfo {author} {\bibfnamefont {T.}~\bibnamefont
  {Hirano}}, \bibinfo {author} {\bibfnamefont {K.}~\bibnamefont {Morita}},
  \bibinfo {author} {\bibfnamefont {S.}~\bibnamefont {Muroya}}, \ and\ \bibinfo
  {author} {\bibfnamefont {C.}~\bibnamefont {Nonaka}},\ }\href {\doibase
  10.1103/PhysRevC.65.061902} {\bibfield  {journal} {\bibinfo  {journal}
  {Phys.Rev.}\ }\textbf {\bibinfo {volume} {C65}},\ \bibinfo {pages} {061902}
  (\bibinfo {year} {2002})},\ \Eprint {http://arxiv.org/abs/nucl-th/0110009}
  {arXiv:nucl-th/0110009 [nucl-th]} \BibitemShut {NoStop}%
\bibitem [{\citenamefont {Schenke}\ \emph {et~al.}(2010)\citenamefont
  {Schenke}, \citenamefont {Jeon},\ and\ \citenamefont
  {Gale}}]{Schenke:2010nt}%
  \BibitemOpen
  \bibfield  {author} {\bibinfo {author} {\bibfnamefont {B.}~\bibnamefont
  {Schenke}}, \bibinfo {author} {\bibfnamefont {S.}~\bibnamefont {Jeon}}, \
  and\ \bibinfo {author} {\bibfnamefont {C.}~\bibnamefont {Gale}},\ }\href
  {\doibase 10.1103/PhysRevC.82.014903} {\bibfield  {journal} {\bibinfo
  {journal} {Phys.Rev.}\ }\textbf {\bibinfo {volume} {C82}},\ \bibinfo {pages}
  {014903} (\bibinfo {year} {2010})},\ \Eprint {http://arxiv.org/abs/1004.1408}
  {arXiv:1004.1408 [hep-ph]} \BibitemShut {NoStop}%
\bibitem [{\citenamefont {Wu}(1997)}]{wu1997maximum}%
  \BibitemOpen
  \bibfield  {author} {\bibinfo {author} {\bibfnamefont {N.}~\bibnamefont
  {Wu}},\ }\href {http://books.google.com/books?id=y4R0QgAACAAJ} {\emph
  {\bibinfo {title} {The Maximum Entropy Method}}},\ Data and Knowledge in a
  Changing World\ (\bibinfo  {publisher} {Springer-Verlag},\ \bibinfo {year}
  {1997})\BibitemShut {NoStop}%
\bibitem [{\citenamefont {Bryan}(1990)}]{bryan}%
  \BibitemOpen
  \bibfield  {author} {\bibinfo {author} {\bibfnamefont {R.}~\bibnamefont
  {Bryan}},\ }\href {\doibase 10.1007/BF02427376} {\bibfield  {journal}
  {\bibinfo  {journal} {European Biophysics Journal}\ }\textbf {\bibinfo
  {volume} {18}},\ \bibinfo {pages} {165} (\bibinfo {year} {1990})}\BibitemShut
  {NoStop}%
\bibitem [{\citenamefont {Cooper}\ and\ \citenamefont
  {Frye}(1974)}]{Cooper:1974mv}%
  \BibitemOpen
  \bibfield  {author} {\bibinfo {author} {\bibfnamefont {F.}~\bibnamefont
  {Cooper}}\ and\ \bibinfo {author} {\bibfnamefont {G.}~\bibnamefont {Frye}},\
  }\href {\doibase 10.1103/PhysRevD.10.186} {\bibfield  {journal} {\bibinfo
  {journal} {Phys.Rev.}\ }\textbf {\bibinfo {volume} {D10}},\ \bibinfo {pages}
  {186} (\bibinfo {year} {1974})}\BibitemShut {NoStop}%
\bibitem [{\citenamefont {Jarrell}\ and\ \citenamefont
  {Gubernatis}(1996)}]{Jarrell1996133}%
  \BibitemOpen
  \bibfield  {author} {\bibinfo {author} {\bibfnamefont {M.}~\bibnamefont
  {Jarrell}}\ and\ \bibinfo {author} {\bibfnamefont {J.}~\bibnamefont
  {Gubernatis}},\ }\href {\doibase
  http://dx.doi.org/10.1016/0370-1573(95)00074-7} {\bibfield  {journal}
  {\bibinfo  {journal} {Physics Reports}\ }\textbf {\bibinfo {volume} {269}},\
  \bibinfo {pages} {133 } (\bibinfo {year} {1996})}\BibitemShut {NoStop}%
\bibitem [{\citenamefont {Asakawa}\ \emph {et~al.}(2001)\citenamefont
  {Asakawa}, \citenamefont {Hatsuda},\ and\ \citenamefont
  {Nakahara}}]{Asakawa:2000tr}%
  \BibitemOpen
  \bibfield  {author} {\bibinfo {author} {\bibfnamefont {M.}~\bibnamefont
  {Asakawa}}, \bibinfo {author} {\bibfnamefont {T.}~\bibnamefont {Hatsuda}}, \
  and\ \bibinfo {author} {\bibfnamefont {Y.}~\bibnamefont {Nakahara}},\ }\href
  {\doibase 10.1016/S0146-6410(01)00150-8} {\bibfield  {journal} {\bibinfo
  {journal} {Prog.Part.Nucl.Phys.}\ }\textbf {\bibinfo {volume} {46}},\
  \bibinfo {pages} {459} (\bibinfo {year} {2001})},\ \Eprint
  {http://arxiv.org/abs/hep-lat/0011040} {arXiv:hep-lat/0011040 [hep-lat]}
  \BibitemShut {NoStop}%
\bibitem [{\citenamefont {{Bajkova}}(1992)}]{Bajkova:1992}%
  \BibitemOpen
  \bibfield  {author} {\bibinfo {author} {\bibfnamefont {A.~T.}\ \bibnamefont
  {{Bajkova}}},\ }\href {\doibase 10.1080/10556799208230532} {\bibfield
  {journal} {\bibinfo  {journal} {Astronomical and Astrophysical Transactions}\
  }\textbf {\bibinfo {volume} {1}},\ \bibinfo {pages} {313} (\bibinfo {year}
  {1992})}\BibitemShut {NoStop}%
\bibitem [{\citenamefont {Ding}(2010)}]{ding}%
  \BibitemOpen
  \bibfield  {author} {\bibinfo {author} {\bibfnamefont {H.-T.}\ \bibnamefont
  {Ding}},\ }\emph {\bibinfo {title} {Charmonium correlation and spectral
  functions in quenched lattice QCD at finite temperature}},\ \href@noop {}
  {Ph.D. thesis},\ \bibinfo  {school} {Bielefeld University} (\bibinfo {year}
  {2010})\BibitemShut {NoStop}%
\bibitem [{\citenamefont {Khalatnikov}(1954)}]{kh:1954}%
  \BibitemOpen
  \bibfield  {author} {\bibinfo {author} {\bibfnamefont {I.}~\bibnamefont
  {Khalatnikov}},\ }\href@noop {} {\bibfield  {journal} {\bibinfo  {journal}
  {Zh.Eksp. Teor. Fiz.}\ }\textbf {\bibinfo {volume} {36}},\ \bibinfo {pages}
  {529} (\bibinfo {year} {1954})}\BibitemShut {NoStop}%
\bibitem [{\citenamefont {Belenkij}\ and\ \citenamefont
  {Landau}(1956)}]{Belenkij:1956cd}%
  \BibitemOpen
  \bibfield  {author} {\bibinfo {author} {\bibfnamefont {S.}~\bibnamefont
  {Belenkij}}\ and\ \bibinfo {author} {\bibfnamefont {L.}~\bibnamefont
  {Landau}},\ }\href@noop {} {\bibfield  {journal} {\bibinfo  {journal} {Nuovo
  Cim.Suppl.}\ }\textbf {\bibinfo {volume} {3S10}},\ \bibinfo {pages} {15}
  (\bibinfo {year} {1956})}\BibitemShut {NoStop}%
\bibitem [{\citenamefont {Landau}\ and\ \citenamefont
  {Lifshit︠s︡}(1959)}]{landau1959fluid}%
  \BibitemOpen
  \bibfield  {author} {\bibinfo {author} {\bibfnamefont {L.}~\bibnamefont
  {Landau}}\ and\ \bibinfo {author} {\bibfnamefont {E.}~\bibnamefont
  {Lifshit︠s︡}},\ }\href {http://books.google.com/books?id=v6kNAQAAIAAJ}
  {\emph {\bibinfo {title} {Fluid mechanics}}},\ A-W series in advanced
  physics\ (\bibinfo  {publisher} {Pergamon Press},\ \bibinfo {year}
  {1959})\BibitemShut {NoStop}%
\bibitem [{Note1()}]{Note1}%
  \BibitemOpen
  \bibinfo {note} {Equation.~\protect \textup {\hbox {\mathsurround \z@
  \protect \normalfont (\ignorespaces \ref {eq:kheq}\unskip \@@italiccorr )}}
  has been applied to study longitudinal expansion in heavy-ion collisions in
  Ref.~\cite {Bialas:2007iu,*Beuf:2008vd,*Peschanski:2010cs}
  recently.}\BibitemShut {Stop}%
\bibitem [{\citenamefont {Lappi}\ and\ \citenamefont
  {McLerran}(2006)}]{Lappi:2006fp}%
  \BibitemOpen
  \bibfield  {author} {\bibinfo {author} {\bibfnamefont {T.}~\bibnamefont
  {Lappi}}\ and\ \bibinfo {author} {\bibfnamefont {L.}~\bibnamefont
  {McLerran}},\ }\href {\doibase 10.1016/j.nuclphysa.2006.04.001} {\bibfield
  {journal} {\bibinfo  {journal} {Nucl.Phys.}\ }\textbf {\bibinfo {volume}
  {A772}},\ \bibinfo {pages} {200} (\bibinfo {year} {2006})},\ \Eprint
  {http://arxiv.org/abs/hep-ph/0602189} {arXiv:hep-ph/0602189 [hep-ph]}
  \BibitemShut {NoStop}%
\bibitem [{\citenamefont {Grumiller}\ and\ \citenamefont
  {Romatschke}(2008)}]{Grumiller:2008va}%
  \BibitemOpen
  \bibfield  {author} {\bibinfo {author} {\bibfnamefont {D.}~\bibnamefont
  {Grumiller}}\ and\ \bibinfo {author} {\bibfnamefont {P.}~\bibnamefont
  {Romatschke}},\ }\href {\doibase 10.1088/1126-6708/2008/08/027} {\bibfield
  {journal} {\bibinfo  {journal} {JHEP}\ }\textbf {\bibinfo {volume} {0808}},\
  \bibinfo {pages} {027} (\bibinfo {year} {2008})},\ \Eprint
  {http://arxiv.org/abs/0803.3226} {arXiv:0803.3226 [hep-th]} \BibitemShut
  {NoStop}%
\bibitem [{\citenamefont {Gubser}(2013)}]{Gubser:2012gy}%
  \BibitemOpen
  \bibfield  {author} {\bibinfo {author} {\bibfnamefont {S.~S.}\ \bibnamefont
  {Gubser}},\ }\href {\doibase 10.1103/PhysRevC.87.014909} {\bibfield
  {journal} {\bibinfo  {journal} {Phys.Rev.}\ }\textbf {\bibinfo {volume}
  {C87}},\ \bibinfo {pages} {014909} (\bibinfo {year} {2013})},\ \Eprint
  {http://arxiv.org/abs/1210.4181} {arXiv:1210.4181 [hep-th]} \BibitemShut
  {NoStop}%
\bibitem [{\citenamefont {Chesler}\ and\ \citenamefont
  {Yaffe}(2011)}]{Chesler:2010bi}%
  \BibitemOpen
  \bibfield  {author} {\bibinfo {author} {\bibfnamefont {P.~M.}\ \bibnamefont
  {Chesler}}\ and\ \bibinfo {author} {\bibfnamefont {L.~G.}\ \bibnamefont
  {Yaffe}},\ }\href {\doibase 10.1103/PhysRevLett.106.021601} {\bibfield
  {journal} {\bibinfo  {journal} {Phys.Rev.Lett.}\ }\textbf {\bibinfo {volume}
  {106}},\ \bibinfo {pages} {021601} (\bibinfo {year} {2011})},\ \Eprint
  {http://arxiv.org/abs/1011.3562} {arXiv:1011.3562 [hep-th]} \BibitemShut
  {NoStop}%
\bibitem [{\citenamefont {Casalderrey-Solana}\ \emph
  {et~al.}(2013)\citenamefont {Casalderrey-Solana}, \citenamefont {Heller},
  \citenamefont {Mateos},\ and\ \citenamefont {van~der
  Schee}}]{Casalderrey-Solana:2013aba}%
  \BibitemOpen
  \bibfield  {author} {\bibinfo {author} {\bibfnamefont {J.}~\bibnamefont
  {Casalderrey-Solana}}, \bibinfo {author} {\bibfnamefont {M.~P.}\ \bibnamefont
  {Heller}}, \bibinfo {author} {\bibfnamefont {D.}~\bibnamefont {Mateos}}, \
  and\ \bibinfo {author} {\bibfnamefont {W.}~\bibnamefont {van~der Schee}},\
  }\href {\doibase 10.1103/PhysRevLett.111.181601} {\bibfield  {journal}
  {\bibinfo  {journal} {Phys. Rev. Lett. 111,}\ }\textbf {\bibinfo {volume}
  {181601}},\ \bibinfo {pages} {181601} (\bibinfo {year} {2013})},\ \Eprint
  {http://arxiv.org/abs/1305.4919} {arXiv:1305.4919 [hep-th]} \BibitemShut
  {NoStop}%
\bibitem [{\citenamefont {Bialas}\ \emph {et~al.}(2007)\citenamefont {Bialas},
  \citenamefont {Janik},\ and\ \citenamefont {Peschanski}}]{Bialas:2007iu}%
  \BibitemOpen
  \bibfield  {author} {\bibinfo {author} {\bibfnamefont {A.}~\bibnamefont
  {Bialas}}, \bibinfo {author} {\bibfnamefont {R.}~\bibnamefont {Janik}}, \
  and\ \bibinfo {author} {\bibfnamefont {R.~B.}\ \bibnamefont {Peschanski}},\
  }\href {\doibase 10.1103/PhysRevC.76.054901} {\bibfield  {journal} {\bibinfo
  {journal} {Phys.Rev.}\ }\textbf {\bibinfo {volume} {C76}},\ \bibinfo {pages}
  {054901} (\bibinfo {year} {2007})},\ \Eprint {http://arxiv.org/abs/0706.2108}
  {arXiv:0706.2108 [nucl-th]} \BibitemShut {NoStop}%
\bibitem [{\citenamefont {Beuf}\ \emph {et~al.}(2008)\citenamefont {Beuf},
  \citenamefont {Peschanski},\ and\ \citenamefont {Saridakis}}]{Beuf:2008vd}%
  \BibitemOpen
  \bibfield  {author} {\bibinfo {author} {\bibfnamefont {G.}~\bibnamefont
  {Beuf}}, \bibinfo {author} {\bibfnamefont {R.}~\bibnamefont {Peschanski}}, \
  and\ \bibinfo {author} {\bibfnamefont {E.~N.}\ \bibnamefont {Saridakis}},\
  }\href {\doibase 10.1103/PhysRevC.78.064909} {\bibfield  {journal} {\bibinfo
  {journal} {Phys.Rev.}\ }\textbf {\bibinfo {volume} {C78}},\ \bibinfo {pages}
  {064909} (\bibinfo {year} {2008})},\ \Eprint {http://arxiv.org/abs/0808.1073}
  {arXiv:0808.1073 [nucl-th]} \BibitemShut {NoStop}%
\bibitem [{\citenamefont {Peschanski}\ and\ \citenamefont
  {Saridakis}(2011)}]{Peschanski:2010cs}%
  \BibitemOpen
  \bibfield  {author} {\bibinfo {author} {\bibfnamefont {R.}~\bibnamefont
  {Peschanski}}\ and\ \bibinfo {author} {\bibfnamefont {E.~N.}\ \bibnamefont
  {Saridakis}},\ }\href {\doibase 10.1016/j.nuclphysa.2010.10.009} {\bibfield
  {journal} {\bibinfo  {journal} {Nucl.Phys.}\ }\textbf {\bibinfo {volume}
  {A849}},\ \bibinfo {pages} {147} (\bibinfo {year} {2011})},\ \Eprint
  {http://arxiv.org/abs/1006.1603} {arXiv:1006.1603 [hep-th]} \BibitemShut
  {NoStop}%
\bibitem [{\citenamefont {Hobson}\ and\ \citenamefont
  {Lasenby}(1998)}]{Hobson:1998bz}%
  \BibitemOpen
  \bibfield  {author} {\bibinfo {author} {\bibfnamefont {M.}~\bibnamefont
  {Hobson}}\ and\ \bibinfo {author} {\bibfnamefont {A.}~\bibnamefont
  {Lasenby}},\ }\href@noop {} {\  (\bibinfo {year} {1998})},\ \Eprint
  {http://arxiv.org/abs/astro-ph/9810240} {arXiv:astro-ph/9810240 [astro-ph]}
  \BibitemShut {NoStop}%
\bibitem [{\citenamefont {Jeffreys}(1998)}]{jeffreys1998theory}%
  \BibitemOpen
  \bibfield  {author} {\bibinfo {author} {\bibfnamefont {H.}~\bibnamefont
  {Jeffreys}},\ }\href {http://books.google.com/books?id=vh9Act9rtzQC} {\emph
  {\bibinfo {title} {The Theory of Probability}}}\ (\bibinfo  {publisher} {OUP
  Oxford},\ \bibinfo {year} {1998})\BibitemShut {NoStop}%
\bibitem [{Note2()}]{Note2}%
  \BibitemOpen
  \bibinfo {note} {The output of MEM typically contains wiggles -- a known
  artifact of the method~\cite {ding}. We applied a moving average filter to
  suppress them.}\BibitemShut {Stop}%
\end{thebibliography}%

\begin{appendix}
\section{A brief review of {the} maximum entropy method}
\label{sec:MEM_review}
In this section, we briefly review essential details of the maximum
entropy method (MEM) used in the present paper to reconstruct the
freeze-out surface.  For further details and explanation an
interested reader may refer to review\cite{Asakawa:2000tr} or
textbook\cite{wu1997maximum}.

Let us begin by putting the Cooper-Frye integral transformation
\eq\eqref{eq:Cooper-Frye-2dv2} into the following form:
\begin{equation}
n_{\th}(Y) =
\int^{\infty}_{0} d\a\, 
\[\,K_1(Y;\a)\rho_1(\a)\, + K_2(Y;\a)\rho_2(\a)\,\]\, ,
\end{equation}
where $n_{\th}(Y)\equiv (dN/dY)_{\th}$ denotes the theoretically
expected particle rapidity distribution given freeze-out surface size,
shape and flow encoded in functions  $\rho_{1,2}(\a)$
defined in Eq.~(\ref{eq:rho12_def}). Here, \bes
\begin{multline}
K_1(Y;\a) =
\frac{ \mt^2}{\phase\Tf}\Big[\,
e^{-\frac{\mt}{\Tf}\cosh(Y-\a)} \\\times
\sinh^2(Y-\a)
+ (\a\to-\a)
\,\Big]\, ;
\end{multline}
\begin{multline}
K_2(Y;\a)=
\frac{\mt^2}{\phase\Tf}\Big[\,
e^{-\frac{\mt}{\Tf} \cosh(Y-\a)}\\\times
\sinh(Y-\a)\cosh(Y-\a)
- (\a\to-\a)
\,\Big]\, .
\end{multline}
\ees
are kernels directly determined from \eq\eqref{eq:Cooper-Frye-2dv2} and we used
$\rho_1(\a)=\rho_1(-\a)$ and $\rho_2(\a)=-\rho_2(-\a)$.

The standard $\chi^2$ fit to reconstruct the ``image'' functions $\rho_1(\a),\rho_2(\a)$ from the data $n_{exp}$ amounts to minimizing the usual likelihood functional:
\begin{multline}
  E(\rho_1,\rho_2) \equiv \frac{1}{2}\sum_{Y, Y'} C^{-1}(Y,Y')
  \[
  n_{\th}(Y)-n_{\exp}(Y)
  \]\\\times
  \[
  n_{\th}(Y')-n_{\exp}(Y')
  \]\, .
\end{multline}
Here,
\begin{equation}
C(Y,Y')\equiv 
\<\, \[ n_{\th}(Y)-n_{\exp}(Y)\]
\[ n_{\th}(Y')-n_{\exp}(Y')\]\,\>
\end{equation}
is the ``covariance matrix" {characterizing expected deviations
between the theoretical model and the experimental results}.
Typically
the number of the data points is smaller 
than that needed to adequately characterize $\rho_1(\a),\rho_2(\a)$.
As a result, there are infinitely many minima of the likelihood
functional, or ``energy", $E(\rho_{1},\rho_{2})$ in $\rho_{1},\rho_{2}$
space -- there are flat directions in that space.
A sensible question one may ask in order to lift this degeneracy would be: what
is the probability distribution of $\rho_{1,2}$ given the data as well
as our prior estimation 
$m_{1},m_{2}$ of $\rho_{1}, \rho_{2}$? We can express this probability as
\begin{equation}
\label{eq:Probability}
P_{\b}(\rho_{1}, \rho_{2}|m_{1},m_{2})
= e^{-\beta F_{\b}(\rho_{1},\rho_{2};m_{1},m_{2})}\, .
\end{equation}
Here, the ``free energy" contains the likelihood functional $E$ and the entropy functional $S$:
\begin{equation}
\beta F_{\beta} 
 = S(\rho_1, \rho_2;m_1,m_2) -\beta E(\rho_1,\rho_2)\, .
\end{equation}
The ``inverse temperature" $\b$ here will balance the relative
importance between the data (given by the ``energy term" $E$) and our
prior estimate (``entropy term" $S$).

Assuming no correlation between $\rho_1,\rho_2$,
we can write $S(\rho_1,\rho_2)$ as
\begin{equation}
\label{eq:S12}
S(\rho_1,\rho_2)
= S_1(\rho_1)+S_2(\rho_2)\, .
\end{equation}
For the case at hand where $\rho_1(\a)= A_\perp
\tau_f(\a)u^{\tau}_f(\a)$ is positive definite and the sign of
$\rho_2(\a)=A_\perp\tau_f(\a)u^{\eta}_f(\a)$ can, in principle, be
either positive or negative, we have: \bes
\label{eq:S1S2}
\begin{equation}
\label{eq:S1}
S_1(\rho_1)=
\int^{\infty}_{0} d\a
 \, \[\,\rho_1(\a)-m_1(\a)-\rho_1(\a)
 \log\frac{\rho_1(\a)}{m_1(\a)}\,\]\, ,
\end{equation}
\begin{multline}
\label{eq:S2}
		S_2(\rho_2)=
\int^{\infty}_{0} 
d\a
 \, \Bigg[\,\sqrt{\rho^2_2(\a)+4m^2_2(\a)} -2m_2(\a) \\
 -\rho_2(\a)\log\frac{\sqrt{\rho^2_2(\a)+4m^2_2(\a)}+\rho_2(\a)}{2m_2(\a)}\,\Bigg]\,
 .
\end{multline}
\ees
Both expressions for the entropy \eq\eqref{eq:S1} and \eq\eqref{eq:S2}
are derived using the law of large numbers (see Ref.~\cite{Asakawa:2000tr} for example for the derivation of \eq\eqref{eq:S1} and Ref.~\cite{Hobson:1998bz} for \eq\eqref{eq:S2}). 
Eq.~(\ref{eq:S1}) is the standard Shannon-Jaynes entropy used for reconstructing arbitrary positive function
(e.g., spectral density in lattice applications~\citep{Asakawa:2000tr}) and
\eq\eqref{eq:S2} is the extended version of the Shannon-Jaynes entropy used for reconstructing image function whose sign is indefinite\cite{Hobson:1998bz,ding}.
The form of $S(F_1,F_2)$ adapted here,
i.e., Eqs.~\eqref{eq:S12} has been applied to deconvolute {\em complex} image functions previously\cite{Bajkova:1992}.

Given the probability distribution in Eq.~(\ref{eq:Probability}) we
can determine an expectation value as a ``weighted average''
of the
image functions $\rho_{1,2}$:
\begin{multline}
\label{eq:FMEM}
  \rho^{\MEM}_{1,2}(\a) = \langle\,\rho_{1,2}(\a)\,\rangle\equiv
 Z^{-1} \int \frac{d\b}{\b} \int {\cal D}\rho_{1}(\a)
{\cal D}\rho_2(\a)\\\times\, 
e^{-\beta F_{\b}(\rho_{1},\rho_{2};m_{1},m_{2})}\, \rho_{1,2}(\a)\, ,
\end{multline}
where, rather than picking a particular value for $\beta$, we followed
a commonly used  Jeffreys' rule\cite{jeffreys1998theory} and
integrated over a scale-invariant measure $d\beta/\beta$. The
normalization constant $Z$ is fixed by requiring $\langle 1\rangle =1$.

\section{Maximum entropy reconstruction of the freeze-out surface.}
\label{sec:MEM_detail}

In practice, 
the functional integral in \eq\eqref{eq:FMEM}
is evaluated in the saddle point approximation.
The saddle point $\rho_1,\rho_2$ is determined  by minimizing $\beta F_{\beta}(\rho_1,\rho_2)$:
\begin{equation}
\label{eq:QF12}
\frac{\delta F_{\b}(\rho_1,\rho_2)}{\delta \rho_1}
=0\, ,
\qquad
\frac{\delta F_{\b}(\rho_1,\rho_2)}{\delta \rho_2}
=0\, .
\end{equation}
As one can show,
for example along the lines of Ref.~\cite{Asakawa:2000tr},
the solution to \eq\eqref{eq:QF12} is unique if it exists.
{The contribution of configurations close to the saddle point 
is included by approximating  $F_\beta$ by a Gaussian.}
We have developed a Mathematica package incorporating Bryan's algorithm\cite{bryan} to find 
$\rho_{1,2}(\a)$  minimizing $\beta F_{\beta}(\rho_1,\rho_2)$
and to evaluate $\rho^{\MEM}_{1,2}(\a)$ as given by
\eq\eqref{eq:FMEM}.
 
As explained in Sec.~\ref{sec:surface},
rapidity-dependent pion distribution is taken from Au-Au collision
data at $\sqrt s=200$~GeV\citep{Bearden:2004yx}.
For simplicity,
we assume the covariance matrix is diagonal
with relative errors of $3$ percent
that 
$C(Y,Y') \approx\(0.03 n_{\exp}(Y)\)^2\delta_{Y,Y'}$.
Incorporating a more elaborate covariance matrix is straightforward.
To calculate functional derivative \eq\eqref{eq:QF12} numerically,
we also discretize fluid rapidity space from $0$ to $\a_{\max}=6$ into $60$ equally-spaced pixels with spacing $\Delta\a = 0.1 $. 
Our choice of spacing $\Delta\a = 0.1$ in fluid rapidity space is guided by the actual spacing in the spatial rapidity space used by such hydrodynamic simulations as Ref.~\cite{Schenke:2010nt}.

 \begin{figure}[htb]
  \centering
	\includegraphics[width=22em]{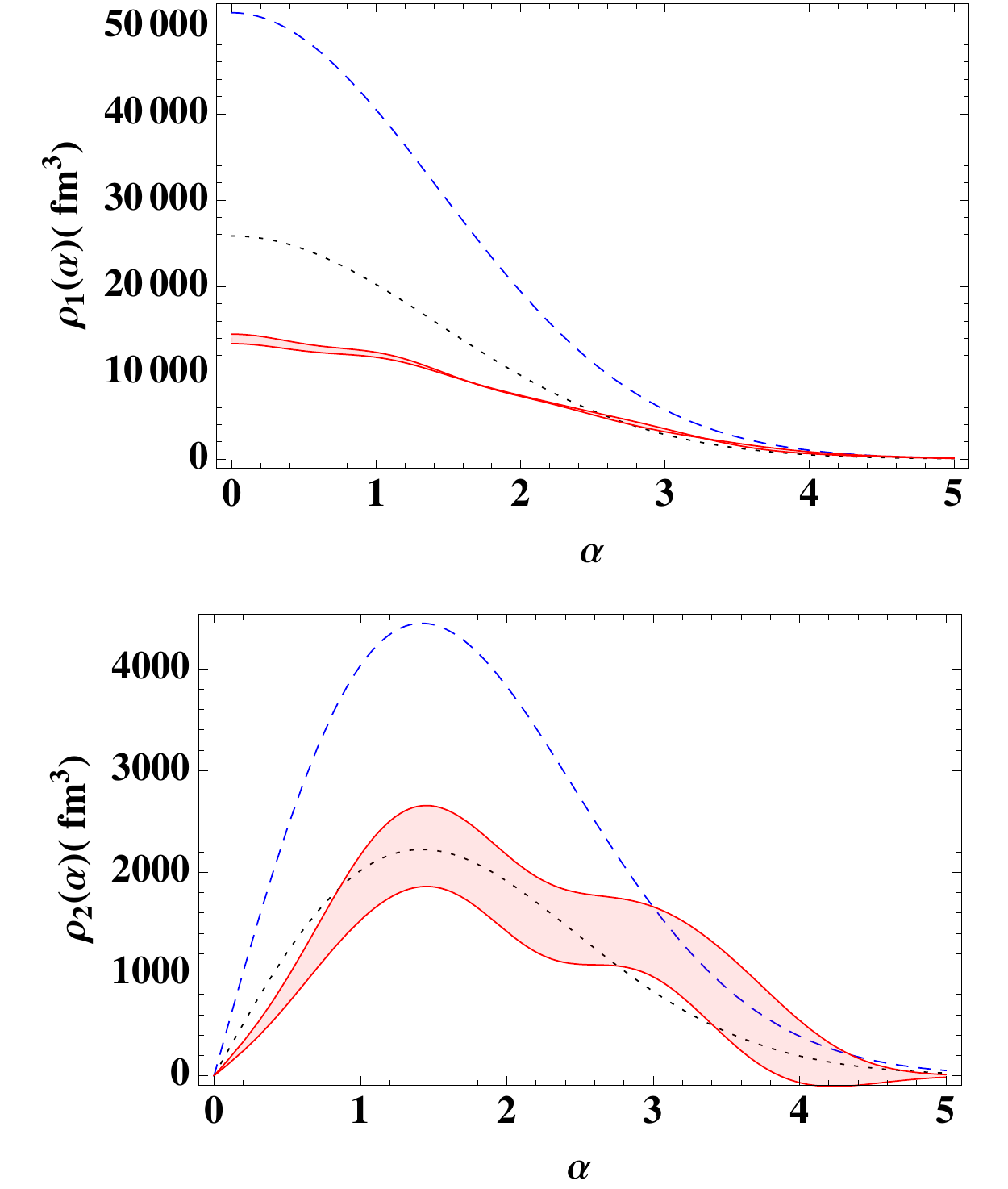}
\caption{
(Color online) The default model dependence of MEM reconstruction.
 Bands shown here are obtained by taking different default models (blue
 and black dashed lines).
(a): $A_{\perp}\tau_f(\a)u^{\tau}_f(\a)$ or $\rho_1(\a)$.
(b): $A_{\perp}\tau^2_f(\a)u^{\eta}_f(\a)$ or $\rho_{2}(\a)$.
}
\label{fig:MEM}
\end{figure}
Motivated by the results of previous hydrodynamic simulations\cite{Satarov:2006iw,Bozek:2009ty},
we parameterize our prior estimate of $\tau_f(\a), \a_f(\a)$ by three parameters $a,b,c$
\begin{equation}
\label{eq:default}
A_{\perp}\tau^{\prior}(\a)
= a\, e^{-b\,\a^2}\, ,
\qquad 
\a - \eta^{\prior}(\a)= c\,\a
\end{equation}
where $\tau^{\prior},\eta^{\prior}$ 
are related to default models in \eq\eqref{eq:S1S2} by
\bes
\begin{align}
&m_1(\a)=A_{\perp}\tau^{\prior}(\a)\cosh(\a - \eta^{\prior}(\a));\\
&m_2(\a)=A_{\perp}\tau^{\prior}(\a)\sinh(\a-\eta^{\prior}(\a)). 
\end{align}
\ees

We have checked the default model dependence of MEM reconstruction with  various choices of $(a,b,c)$. 
To provide a transparent idea of the sensitivity of our results to the choice of default model,
in Fig.~\ref{fig:MEM},
we have plotted the output of the maximum entropy method $\rho_{1}(\alpha), \rho_{2}(\alpha)$
for  two different choices of $a$ and fixed $b,c$.
\footnote{
The output of MEM typically contains wiggles -- a known artifact of
the method~\cite{ding}. We applied a moving average filter to suppress
them.}. 
The error band in Fig.~\ref{fig:equaltau} is based on results of our analysis using those two default models. 
From Fig.~\ref{fig:MEM},
we notice that $\rho_{1}(\a)$ reconstructed using the maximum
entropy method is relatively insensitive to the choice of the default
model. This can be understood as a consequence of the fact that the
error bars on the data are small and about $80-90\%$ of
the contribution to $dN/dY$ comes from $\rho_1(\a)$.
On the other hand,
since the contribution of $\rho_2(\a)$ to $dN/dY$ is much smaller, the
data constrains  $\rho_2(\a)$ much less, and thus the sensitivity to the
choice of the default model is stronger, as seen in Fig.~\ref{fig:MEM}.

\end{appendix}


\end{document}